\documentclass[english]{article}
\usepackage{lmodern}

\usepackage[T1]{fontenc}
\usepackage[latin9]{inputenc}
\usepackage{babel}
\usepackage{listings}
\usepackage[unicode=true,
 bookmarks=true,bookmarksnumbered=false,bookmarksopen=false,
 breaklinks=true,pdfborder={0 0 0},backref=false,colorlinks=false]
 {hyperref}
\hypersetup{pdftitle={Finitary-based Domain Theory in Coq: An Early Report},
 pdfauthor={Moez A. AbdelGawad},
 pdfsubject={Domain Theory in Coq},
 pdfkeywords={Domain Theory; Coq; Finitary-basis}}

\makeatletter
\newenvironment{lyxcode}
{\par\begin{list}{}{
\setlength{\rightmargin}{\leftmargin}
\setlength{\listparindent}{0pt}
\raggedright
\setlength{\itemsep}{0pt}
\setlength{\parsep}{0pt}
\normalfont\ttfamily}%
\item[]}
{\end{list}}

\newcommand{\code}[1]{\texttt{#1}}

\makeatother

\begin{document}

\title{Finitary-based Domain Theory in Coq:\\
An Early Report\footnote{This article is based on an earlier extended abstract that was published in Coq7. The extended abstract did not include code examples.}}

\author{Moez A. AbdelGawad\\
Informatics Research Institute, SRTA-City\\
Alexandria, Egypt\\
\texttt{moez@cs.rice.edu}}

\maketitle
\begin{abstract}
	In domain theory every finite computable object can be represented by a single mathematical object instead of a set of objects, using the notion of finitary basis. In this article we report on our effort to formalize domain theory in Coq in terms of finitary basis.
\end{abstract}
\section{Introduction}
In his ``Lectures on a Mathematical Theory of Computation''~\cite{ScottCompMathTheoryLects},
Dana Scott formulated domains in terms of neighborhood systems. Later,
Scott favored a formulation in terms of information systems~\cite{Scott82}
but has not rewritten his lectures notes. Cartwright, Parsons and the author later
revised Scott's lecture notes to reflect a formulation of domains
in terms of `finitary basis'~\cite{DomTheoryIntro}, where a finitary
basis is an information system that is closed under least upper bounds
on finite consistent subsets. Finitary basis have the desirable property
that every finite computable object is represented by a single basis
element instead of a set of elements.

In 2009 we started an effort to formalize Cartwright et~al's domain theory monograph in Coq~\cite{Bertot2004}. When we finished formalizing
only the first chapter of the monograph, the size of our development was over 2,500 lines of Coq code. Our goal was to publish the formalization of the full monograph online when the monograph itself gets published
as a primer on domain theory. In this article we discuss how our effort progressed.
\section{Our Formalization}
When we started our formalization of domain theory in Coq we were met with initial ease, but that was followed by a difficulty of proceeding
at the same initial pace.  We believe the initial ease we met in our effort is due to Coq's
``built-in'' support, via its libraries, for set theory and partial orders (posets).
We believe the ``formalization resistance'' we later met as we dug into our formalization of domain
theory, however, is due to the ``thickness'' of layers upon layers of definitions
of domain theory (which is typical of any mathematical discipline),
combined with Coq's unstructured proof syntax and with proof states
being implicit and not explicitly stated in Coq proofs. In our experience,
these factors have made proofs in the domain theory
monograph that are relatively simple\footnote{As presented in~\cite{DomTheoryIntro}, and as their counterparts are presented in Scott's work, these proofs prove facts that are ``obvious'' or almost obvious.} become lengthy and much harder to
grasp when formalized in Coq.

In particular, since finitary basis are restricted partial orders, the initial ease we found in writing domain theory
definitions and constructing proofs in Coq was due to the support
for set theory in Coq (via type \code{Ensemble}) and for order theory/posets
(via type \code{PO}). This made the first few initial definitions
and proofs in our formalization straightforward.

However, as we demonstrate in Section~\ref{sec:code}, we faced hardship later. Examples of proofs we found unnecessarily
hard (i.e., lengthy and time consuming) to develop in Coq are the proofs that
\begin{itemize}
\item the singleton set containing bottom is the bottom (i.e., smallest) ideal,
\item a finite subset of a union set has a finite covering set, and
\item the union of a directed set of ideals is an ideal.
\end{itemize}
which we present, with few comments, below.
\subsection{Coq Code Excerpts}
\label{sec:code}
In this section we present Coq code examples that demonstrate how our formalization of domain theory was met with resistance after initial ease.
\subsubsection{Initial Definitions and Proofs}
\begin{lyxcode}
\begin{lstlisting}[language=ML,tabsize=4]
Section Domains.

Variable U : Type.

Require Import Partial_Order.

(* a partial order quadruple over a subset of U *)
Variable P : PO U.
\end{lstlisting}
\end{lyxcode}

Based on these definitions we proved some initial set of simple theorems.  Most proofs were under fifteen lines of Coq code, as demonstrated by the following proof of the uniqueness of least upper bounds.

\begin{lyxcode}
\begin{lstlisting}
Theorem LubsUnique:
  forall (S1: Ensemble U)(S2: Ensemble U)(b: U)(c: U),
     LubIn S1 S2 b -> LubIn S1 S2 c -> b = c.
Proof.
  intros S1 S2 b c lb lc.
  destruct lb as [usb rbub]. destruct lc as [usc rcub].
  assert (rcb: R c b).
    apply rcub; assumption.
  assert (rbc: R b c).
    apply rbub; assumption.
  apply P_AS.
    assumption.
  assumption.
Qed.
\end{lstlisting}
\end{lyxcode}
However, as we defined finitary basis and progressed towards proofs that are more relevant to domain theory, proofs tended to get longer, as demonstrated by the proof below that a principal ideal is indeed an ideal.

\begin{lyxcode}
\begin{lstlisting}
Definition FinitaryBasis (B: Ensemble U): Prop :=
  Countable B
  /\ Inhabited U B
     (* Guarantee empty set is consistent/bounded *)
  /\ (forall S: Ensemble U,
      Included U S B -> Finite U S 
      -> UpperBoundedIn B S -> HasLubIn B S).

Definition Directed (S: Ensemble U): Prop :=
  forall F, Included U F S -> Finite U F
    -> UpperBoundedIn S F.

  .
  .
  .
  .
  
Definition IdealIn (B: Ensemble U)(I: Ensemble U): Prop :=
  FinitaryBasis B /\ HasItsDownwardClosuresIn B I 
  /\ DirectedSubsetOf B I.
(* A downward-closed directed subset of basis B is an ideal *)

Definition IsPrIdlIn (B: Ensemble U)(I: Ensemble U): Prop :=
  FinitaryBasis B /\ 
    (exists b: U, In U B b /\ I = LowerSetOfIn B b).

Definition PrIdlOfIn (B: Ensemble U)(b: U)
  (FB: FinitaryBasis B)(InBb: In U B b): Ensemble U :=
  LowerSetOfIn B b.

(* The theorem and its proof are 45 lines. *)
(* All pr. ideals are ideals. Not direct from defn. *)
Theorem PrIdlIsIdl: 
	forall B b FB InBb, IdealIn B (PrIdlOfIn B b FB InBb).
Proof.
  intros B b FB InBb.
  unfold IdealIn.
  split.
    assumption.
  split.
    elim (DC_eqv B (PrIdlOfIn B b FB InBb)).
    intros _ DCO.
    apply DCO.

    intros i InPIi i' InBi' Ri'i.
    unfold In.
    unfold PrIdlOfIn.
    split.
      assumption.
    unfold In in InPIi.
    unfold PrIdlOfIn in InPIi.
    unfold LowerSetOfIn in InPIi.
    destruct InPIi as [InBi Rib].
    apply P_Tr with i.
      assumption.
	assumption.

  unfold DirectedSubsetOf.
  split.
    unfold Included.
    unfold In.
    unfold PrIdlOfIn.
    unfold LowerSetOfIn.
    unfold In.
    intros x assump.
    destruct assump as [Bx _].
    assumption.

  intros SS SSII FSS.
  unfold UpperBoundedIn.
  exists b.
  unfold UpperBoundIn.
  split.
    apply bInPrIdlb.
  intros s InSSs.
  unfold Included in SSII.
  unfold In at 2 in SSII.
  unfold PrIdlOfIn in SSII.
  elim (SSII s).
    intros _ Rsb.
    assumption.
  assumption.
Qed.
\end{lstlisting}
\end{lyxcode}

\subsubsection{Formalization Resistance}
The following three theorems and their Coq proofs below demonstrate the resistance we met close to the end of our formalization.

\paragraph*{$\bullet$ The Singleton Set Containing Bottom is The Bottom Ideal}
\begin{lyxcode}
\begin{lstlisting}[language=ML,tabsize=4]
(* The theorem and its proof are 103 lines. *) 
Theorem BotSingletonIsBotIdeal:
  forall B: Ensemble U, FinitaryBasis B ->
    (exists bot: U, IsBotOf B bot /\ IdealIn B (Singleton U bot) /\
    (forall I: 
       Ensemble U, IdealIn B I -> Included U (Singleton U bot) I)).
Proof.
  intros B FB.
  assert (HasBot B) as BBot.
    apply FBBot.
    assumption.
  unfold HasBot in BBot.
  destruct BBot as [bot botIsBot].
  
  exists bot.
  
  split.
  assumption.
  
  unfold IsBotOf in botIsBot.
  unfold LubIn in botIsBot.
  destruct botIsBot as [botUBEmp RbotB].
  
  split.
    unfold IdealIn.
    split.
      assumption.
    split.
      elim (DC_eqv B (Singleton U bot)).
      intros _ DCO.
      apply DCO.
  
      unfold DownwardClosedIn_orig.
      unfold In.
      intros e eBot b InBb Rbe.
      Require Import Constructive_sets.
      apply Singleton_intro.
      assert (bot=e) as eIsBot.
        apply Singleton_inv.
        unfold In.
        assumption.
      assert (R bot b) as Rbotb.
        apply (RbotB b).
        unfold UpperBoundIn.
        split.
          assumption.
        contradiction.
      rewrite <- eIsBot in Rbe.
      apply P_AS.
        assumption.
      assumption.
  
    unfold DirectedSubsetOf.
    split.
      unfold UpperBoundIn in botUBEmp.
      destruct botUBEmp as [InBbot _].
      unfold Included.
      intros x xbot.
      apply Singleton_inv in xbot.
      rewrite <- xbot.
      assumption.

    intros SS SSbot FSS.
    unfold Included in SSbot.
    unfold UpperBoundedIn.
    exists bot.
    unfold UpperBoundIn.
    split.
      unfold In.
      apply Singleton_intro.
      trivial.
    intros s InSSs.
    elim (SSbot s).
      cut (forall e: U, R e e).
        auto.
      apply P_Rx.
    assumption.
  
  intros I.
  unfold IdealIn.
  elim (DC_eqv B I).
  intros DCO _.
  intros [_ [DCI DI]].
  apply DCO in DCI.
  
  unfold Included.
  intros x xbot.
  apply Singleton_inv in xbot.
  rewrite <- xbot.
  unfold In.
  unfold DownwardClosedIn_orig in DCI.
  clear x xbot.
  unfold DirectedSubsetOf in DI.
  destruct DI as [IinB DI].
  elim (DirInh I).
  intros x InIx.
  apply (DCI x).
      assumption.
    destruct botUBEmp as [InBbot _].
    assumption.
  apply RbotB.
  apply (AllS_UB_Empty B).
  apply IinB.
  assumption.
  assumption.
Qed.

 .
 .
 .
 
End Domains.
\end{lstlisting}
\end{lyxcode}

\paragraph{$\bullet$ A Subset of a Finite Union has a Finite Covering Set}
\begin{lyxcode}
\begin{lstlisting}[language=ML,tabsize=4]
Section Union.

Variable U: Type.

Let S_U := Ensemble U : Type.

Let S_S_U := Ensemble S_U : Type.

 .
 .
 .
 
(* The theorem and its proof are 81 lines. 
   This theorem is proven to shorten the proof of
     UnionOfDirectedIdealsIsAnIdeal below. *)
Theorem FUnionSubsetHasFCoverSet:
        forall (SSU: S_S_U)(SU: S_U),
          Finite U SU -> Included U SU (Union SSU) ->
          (exists SSUS: S_S_U, 
             Finite S_U SSUS /\ Included S_U SSUS SSU
             /\ Included U SU (Union SSUS)).
Proof.
  intros SSU SU FSU SUss.
  induction FSU.
    exists (Empty_set S_U).
    split.
      apply (Empty_is_finite S_U).
    split.
      unfold Included.
      unfold In.
      intros.
      contradiction.
    intros.
    elim (CondsEqv (Empty_set S_U) (Empty_set U)).
    intros _ EqvIf.
    apply EqvIf.
    contradiction.

  destruct IHFSU as [SSUS assump].
    unfold Add in SUss.
    apply InclUn in SUss.
    destruct SUss.
    assumption.
        
  destruct assump as [FSSUS [SSUSss sassump]].
        
  unfold Add in SUss.
  apply InclUn in SUss.
  destruct SUss as [Ass xSss].
  unfold Included, In, Union in xSss.
  destruct (xSss x) as [Unn assump].
    tauto.
  destruct assump as [UnnInSSU xInUnn].
        
  exists (Add S_U SSUS Unn).  (* crucial step *)
  split.
    Require Import Finite_sets_facts.
    apply Add_preserves_Finite.
    assumption.
        
  split.
    unfold Included.
    intros AugUnn AugUnnInSSUS.
    destruct AugUnnInSSUS.
      unfold Included in SSUSss.
      apply (SSUSss x0).
      assumption.
    unfold In in H0.
    destruct H0.
    assumption.

  elim (CondsEqv (Add S_U SSUS Unn) (Add U A x)).
  intros _ EqvIf.
  apply EqvIf.
  intros s sInAx.
  destruct sInAx.
    assert
      (exists SSUSUx0: S_U, 
         In U SSUSUx0 x0 /\ In S_U SSUS SSUSUx0).
      elim (CondsEqv SSUS A).
      intros EqvOnlyIf _.
      apply (EqvOnlyIf sassump).
      assumption.
    destruct H1.
    exists x1.
    destruct H1.
    split.
      assumption.
    unfold In.
    left.
    assumption.
  destruct H0.
  exists Unn.
  split.
    assumption.
  unfold In.
  right.
  unfold In.
  tauto.
Qed.	
\end{lstlisting}
\end{lyxcode}
\paragraph{$\bullet$ The Union of Directed Ideals is an Ideal}
\begin{lyxcode}
\begin{lstlisting}[language=ML,tabsize=4]
Variable P : PO U.

Let B := Carrier_of U P.
Let R := Rel_of U P.

Variable FB : FinitaryBasis U P B.

Let D : PO S_U := (Domain U P B FB).
Let C : S_S_U := Carrier_of S_U D.
(* Is the same as the set of ideals over the FB B *)

Theorem C_Idls: C = (IdealIn U P B).
Proof.
  subst C.
  unfold D.
  unfold Domain.
  unfold Carrier_of.
  trivial.
Qed.

(* The theorem and its proof are 93 lines.
   To shorten the proof, the proof above of
     FUnionSubsetHasFCoverSet is used. *)
Theorem UnionOfDirectedIdealsIsAnIdeal:
  forall (SI: S_S_U), DirectedSubsetOf S_U D C SI
     -> IdealIn U P B (Union SI).
Proof.
  intros SI SID.
  unfold DirectedSubsetOf in SID.
  destruct SID as [SII SID].
  unfold Included in SII.
  unfold In in SII.
        
  split.
    apply FB.
  split.
    elim (DC_eqv U P B (Union SI)).
    intros _ DCO.
    apply DCO.
    unfold DownwardClosedIn_orig.
    intros e eu b bB be.
    unfold In, Union, UnionCond in eu.
    destruct eu as [SU [SIU SUe]].
    unfold In, Union, UnionCond.
    exists SU.
    split.
      assumption.
    assert (IdealIn U P B SU) as SUI.
      apply (SII SU).
      unfold In in SIU.
      assumption.
    unfold IdealIn in SUI.
    destruct SUI as [_ [DCSU DSU]].
    
    elim (DC_eqv U P B SU).
    intros DCO2 _.
    apply DCO2 in DCSU.
    
    unfold DownwardClosedIn_orig in DCSU.
    apply (DCSU e).
        assumption.
      assumption.
    assumption.
    
  unfold DirectedSubsetOf.
  split.
    unfold Included, In, Union, UnionCond.
    intros x UnS.
    destruct UnS as [SU [SUI xSU]].
    assert (IdealIn U P B SU) as SUId.
      apply (SII SU).
      apply SUI.
    unfold IdealIn in SUId.
    destruct SUId as [_ [DCSU DSU]].
    unfold DirectedSubsetOf in DSU.
    destruct DSU as [SUB _].
    unfold Included in SUB.
    apply (SUB x).
    assumption.
        
  intros SU SUUSI FSU.
        
  destruct (FUnionSubsetHasFCoverSet SI SU FSU SUUSI) 
    as [SSU [FSSU [SSUSI SUSSU]]].
        
  destruct (SID SSU) as [UB SSUUB].
    assumption. assumption.
  unfold UpperBoundIn in SSUUB.
  destruct SSUUB as [UBI UBSSs].
  unfold In in UBI.
  elim (SII UB).
    intros _ UBD.
    destruct UBD as [_ UBD].
    unfold DirectedSubsetOf in UBD.
    destruct UBD as [_ SUUB].
    apply (UpperBoundingIncluded U P SU UB (Union SI)).
      apply (SUUB SU).
        unfold In, Rel_of in UBSSs.
        unfold Included.
        intros x xInSU.
        elim (SUSSU x).
          intros x0 [xInx0 x0InSSU].
          apply (UBSSs x0).
            assumption.
          assumption.
        assumption.
      assumption.
    unfold Included.
    intros x InUBx.
    unfold Union, UnionCond, In.
    exists UB.
    split.
      assumption.
    assumption.
  assumption.
Qed.	
\end{lstlisting}
\end{lyxcode}

As presented, the proofs of the three domain theory theorems above are over seventy-five lines of Coq code, which we believe to be unnecessarily
long and makes the proofs hard to follow, particularly since we believe we have used as much intermediate lemmas as possible as a means
for shortening and structuring these proofs. Additionally, the syntax
of Coq proofs (as an almost linear sequence of commands, each of whose
effects is to change an implicit proof state) did not make the three
Coq proofs of these theorems immediately reveal the main ideas of
the proofs.
\section{Concluding Remarks}
An unfortunate consequence of our experience with Coq is that our domain theory
formalization effort has slowed down.
Even though there
is a chance we may keep using Coq as the tool of choice to express
our formalization (given the time and effort we already invested in
it), but we are also considering switching over to using other proof
assistants (such as Isabelle~\cite{Isabelle2015}, thus restarting
our formalization effort almost from scratch), or even giving up on our
formalization effort altogether. As another consequence of our experience
with Coq, we are also considering the possibility of developing our
own vastly less powerful, yet more user-friendly Proof Designer-based
proof assistant~\cite{AbdelGawad2015c, PD2}.

\bibliographystyle{plain}

\end{document}